# AMITY - A HYBRID MENTAL HEALTH APPLICATION

**Srija Santhanam , Kavipriya P , Balamurugan MS**

School of Electronics Engineering , Vellore Institute of Technology , Chennai , India

**ABSTRACT**.

Wellness in trivial terms combines physical, social, and mental well-being. While mental health is generally neglected, long-term success in a person's life is mostly determined by his psychological health and contentment. For a person in distress, professional mental health services are quite expensive, unpopular, and invite a lot of hesitation. Hence, It would be effective to use an Android application that can offer day-to-day therapeutic assistance, meditation sessions, and guidance since it can cater to a massive community instantly. In this paper, we propose a mobile and web application-AMITY with a chat group and chatbot created using a machine learning approach. We have also built a dataset to train the chatbot model that we propose in this paper. We briefly introduce the dataset and the machine learning model in section 3. In section 4, we include the architecture and the development details of the Hybrid application. Next, we present our results on usability and the efficiency of the idea we propose.

## INTRODUCTION

Emotional well-being stands equal to physical well-being and plays a vital role in creating positive relationships, giving back to the community, and feeling accepted in groups. Positive mental health can help us deal with stress, adapt to changes, make difficult decisions, and manage emotions.

Mental health has become a rising concern over the past decade, especially after a pandemic that scaled worldwide and terrorized lakhs of people in their homes. In 2020, the Centers for Disease Control and Prevention in the United States found that there was a 31% increase in the number of people with mental health disorders from the previous year due to COVID-19. Alarmingly, about 381 suicides happen every day in India according to National Crime Records Bureau(2019).

Around 10% of the population of India in 2015-2016 was found to need Counseling services out of which only 28% received treatment. Also, mental health issues are shunned and ostracized since there is not much awareness propagated. There is an inadequacy in the number of mental health professionals and the percentage of funds allocated by the Government of India is minimal. Consequently, psychological therapy is expensive in India and people in turn solicit help from conventional Institutions such as religious establishments, cultural practices, or immediate friends and family.

The vast majority do not treat mental health problems until their effects become severe. Professional help is not immediate. A significant fraction might also find it difficult to share their problems with another individual face to face. It is said that 1 in 4 people have mental health problems in their lifetime. Supply does not meet the demands here. Hence, a vast platform such as an Android application surely is a benefit. A multitude of mental health applications exist offering breathing and meditation practices, sleep stories, and mindfulness exercises. There is this app called 'WoeBot' offering cognitive behavioral therapy exercises to induce sleep. Another application 'BetterHelp' connects users with therapists online and the MoodFit app tracks the mood of the user on a day-to-day basis with which required exercises are suggested. Generally, people need to talk or let out their emotions to someone on an immediate basis to relieve stress, anxiety, or frustration. The problem is that there would not be many that may listen and empathize. Our solution involves a group chat feature that helps people with similar problems connect, talk and help each other out anytime and anywhere which is lacking in the majority of the apps out there.

Emotional Artificial Intelligence uses machine learning to detect and respond to the emotions of humans with technologies such as Computer Vision and Natural Language Processing. Facial Recognition algorithms are used for detecting expressions, while chatbots enable conversations with users based on their mood that could be tracked from their messages. Hence, helping people who are depressed will be made easier and cheaper with Artificial Intelligence. We include an NLP-based chatbot in our application that offers quick therapeutic assistance to the users based on their input. We also suggest diet and exercise plans as an additional feature. Users would also be able to call therapists if they need a professional to talk to.

## 2.RELATED WORKS

There are a few psychotherapeutic approaches used for mental wellness. Cognitive Behavioral Therapy (CBT). Dialectical Behavior Therapy (DBT), Acceptance and Commitment Therapy (ACT), Psychodynamic Therapy, Interpersonal Therapy (IPT), Eye Movement Desensitization and Reprocessing (EMDR), and Mindfulness-Based Therapies, are some of the major psychotherapeutic approaches used. Among the above therapeutic approaches, Cognitive Behavioral Therapy (CBT) is a psychotherapy strategy that emphasizes how a person's thoughts, emotions, and behaviors interact. It is a method that helps people discover and challenge unfavorable or maladaptive attitudes and behaviors to enhance their emotional and mental well-being. It is goal-oriented and time-limited. Obsessive-compulsive disorder (OCD), post-traumatic stress disorder, depression, anxiety disorders, and other mental health issues are all treated using cognitive behavioral therapy (CBT). It is often given by a mental health professional during individual or group treatment sessions. Chatbots use the same psychotherapy strategy to combat users with mild anxiety and low phase. Basic emotion detection can be done using CBT and since the main focus is on emotions that are of short-term period CBT exactly fits in that scenario. With so many people engaging in social media and other applications, machine learning strategies with the use of keywords could potentially help with mental health variation detection [1]. Chatbot uses the psychotherapy strategy of CBT because the emotions felt by humans in a short period are subjected to a particular object, person, or situation so CBT especially combats such scenarios easily [2].

## 2.1 TECHNOLOGIES FOR EMOTION DETECTION

Depression, anxiety, and low mood could be detected by AI methods like Natural Language Processing, Machine learning Algorithms, and Keyword scraping [5]. This could potentially help medical assistance applications. Research-based on Technologies used for mental health assistance, suggests using AI to design efficient Mobile applications[5]. Mobile and wearable Applications are a massive helping hand during difficult times such as a pandemic.People can easily get mental support at their comfort[6],[10]. Psychological Experiments on an elderly community in China for their general wellness which includes general physical health parameters like temperature, heartbeat, and Mental health Parameters as well have been put into assessment using Smart Phone Application which has shown great success for them [7].

Research conducted on 104 Mental health Mobile Apps in the Play Store which is based on Machine Learning Algorithms and Sentiment Analysis has given positive Results. Content, Privacy, aesthetically Pleasing UI, app stability, and Subscription Cost are major factors influencing mental apps are important findings from the research conducted [11]. The design and the surveillance of a qualified and assured medical expert seem to be major factors in curating these kinds of medical applications. A therapist, a psychological expert in the mental health app will bring confidence and security to the user plus add safety and guidance values to the app [13]. The potential involvement of a Clinician gives much more value to the Mental health app, even during the design development of the application, it helps [14]. Studies made on several mental health based mobile applications suggest that meditation, music, movie, and quotes helped in reducing anxiety,depression and improved overall well-being. The research also suggested improving mental health for a general audience [15]. Research and Studies also show that around 78% of total users use Mental health Mobile Applications in particular due to their easy accessibility feature. The studies further state that doctors have seen the apps creating a magic increase in confidence and have provided support like a Personal Trainer. The App has turned out to be a good companion.[16].

## 2.2 PEER SUPPORT

An enormous number of people go through a low phase with issues such as work stress, issues with family, health problems etcetera. When part of the same linguistic group people tend to discuss similar problems and this may bring out any signs of depression.[7].Peer-led training and employment for mentally depressed people have given excellent results [18]. Peer-led support for individuals with mental health problems has increased, especially in English-speaking Nations. As human beings we show support towards other individuals' aspirations, strengths, and effort.[19]. Students at college who are less likely to develop good social health are at high risk of getting into some kind of mental issue such as depression[20].

A survey conducted on 500 people from a Norwegian Mental Health Online Forum has mentioned that around 75% of the people have found the forum easy to talk about their Mental Health related problems online and a majority have sought empowerment via this [21]. Surveys conducted on 15 Women who have gone through postnatal mental illness and used online forums have felt better outcomes in their mental health in such platforms because they think it provided them with a shared understanding.They also realized that using these online forums also reduced the social stigma around them [22]. People that may have gone through some kind of psychotherapeutic treatment may fail to maintain the results. To see if online social chat groups would provide the necessary help to this, a survey was conducted on about 150 patients who were discharged after their mental recovery. This survey conveys that only 22.2% of them found the online chat group experiences a relapse and a majority of 46.5% have found that these online chat groups are effective in maintaining treatment gains [23]. People can support each other through such online forums, especially for mental health.

Studies based on Mental health help on Social Media in China where people were in intense trauma, the peer group did help, says the studies. It also says the Time zones were one of the barriers that were observed and one more positive factor is the size of the group. Bigger groups around 300 felt majorly beneficial. Ultimately the purpose of setting up such forums did give a hand in utmost unforeseen circumstances [27].

## 2.3 RECOMMENDATIONS

Engaging in daily activities can improve emotional well-being. Studies show us that examining, and recommending personalized activities helps in giving more efficient results [3]. A survey conducted on youths and children based on their lifestyle states that people who are likely to have a sedentary lifestyle are likely to develop Mental Illness[28]. Studies also say that now people are in a positive sign of using chatbots for their Mental health along with the chatbot they are expecting some kind of personalization[31]. Chatbot along with Recommendations could identify the issue much earlier and respective medication can be given[32]. Books can bring value or understanding of what the patients are lacking when in their recovery stage, it is also a kind of relaxation. Music and Movies have also been taken into account bringing the utmost relaxation to any stressed-out individual[35][36][37].

## 2.4 CHATBOT NATURAL LANGUAGE MODEL:

Chatbots are programs developed to simulate conversations with users and they are very popular now. There are different approaches to the creation of chatbots including rule-based, Keyword matching, machine learning, and Hybrid approaches. A rule-based approach works best when the input and the response are known.LSTM is often used for training sequential texts and conversational bots. An LSTM-based model is trained and tested on conversations at the IT helpdesk and on a noisy dataset of movie subtitles[8]. The HRED is a two-layered RNN architecture that is used to process dialogue histories longer than 500 words[9]. This paper summarizes different NLP models with respective datasets. It is found that Tree LSTMs perform better in syntactical interpretation in the analysis of text and BLEU(Bilingual Evaluation Understudy) is a metric used to evaluate dialog systems.[17] .A corpus of 500,000 posts on the KoKo platform is labeled and used for training.GloVe pre-trained word embeddings are used over the text and are sent to the Convolutional neural network followed by a gated recurrent unit. The above studies show us that an LSTM-based Machine Learning Approach would be a suitable model for creating therapy chatbot.

## 3 APPROACH:

### 3.1 REQUIREMENT ANALYSIS:

An adequate insight into conversations in mental health therapy sessions was obtained by observing them. The magnitude of relief obtained by patients depends on the right convincing words picked by the counselor. Due to privacy concerns obtaining mental health therapy data is often difficult. Very few existing datasets containing therapy conversations were analyzed such as the International Survey on Emotion Antecedents and Reactions which contains over 7000 samples and the following are the attributes:
i)Sentence-An expression of the situation.
ii)Category-A classification of the sentence into common emotions such as Joy, anger, fear etcetera.
iii)Gender-Gender of the person who spoke the sentence.
iv)Age- Age of the person who spoke the sentence. A small dataset containing attributes of question and answer patterns.

### 3.2 DATASET CREATION:

A dataset with therapeutic conversations was made with help from a psychological counselor. We created a JSON file with attributes:

i)Tag - A topic that uniquely identifies the type of conversation.
ii)Patterns-A variation of conversational text inputs the user may enter.
iii)Response - Possible choices of response to the user. By using the JSON file format we create organized data that can be easily read and processed by machine learning tools. The dataset contains about 72 tags. Out of which 30 tags are trivial mental health question tags, 10 are greetings and introduction tags and the remaining 32 are mental health descriptive tags. The mental health questions that a user could ask are:
 i)"What treatment options are available?",
 ii)" Where can I find a support group?" etc.
Mental health descriptive tags contain patterns of user's questions on anger management, loneliness, trouble with children, etc.

### 3.3 SYSTEM DEVELOPMENT:

The Hybrid application "AMITY" was developed using the Flutter framework which is used for Mobile app development and web Application. With a single codebase, developers can produce high-performance, aesthetically pleasing, natively built mobile applications for both the iOS and Android platforms using the well-liked cross-platform Flutter framework. With the help of Flutter's extensive library of premade UI widgets, developers can quickly and easily produce attractive, responsive mobile applications. Also, because of its quick development cycle, developers may make adjustments and see the effects right away. Moreover, Flutter offers a wealth of capabilities, like quick reload, which expedites development by enabling developers to immediately view changes they make to the code. Developers may readily obtain assistance and resources for their development projects because of the big and expanding Flutter community. Because Flutter is created and maintained by Google, it gives developers utilizing it for their app development projects a sense of confidence and dependability. Faster and more fluid animations and transitions are possible because of Flutter's high-performance optimized architecture.

The Language used for developing the Application is dart language. Google created the programming language, Dart. It is a class-based, object-oriented programming language that is garbage-collected and can be used to create a wide range of applications, including those for the web, mobile devices, desktop computers, and servers. With a syntax that is recognizable to programmers from other languages like Java, C#, and JavaScript, Dart is made to be simple to learn. With features like just-in-time compilation, ahead-of-time compilation, and tree shaking that help to enhance speed, it is also intended to be quick and effective. The Flutter framework's core language for application development is Dart. Dart's Just-in-Time (JIT) and Ahead-of-Time (AOT) compilation techniques provide quick and effective app performance. With frameworks like Flutter and AngularDart, Dart may be used to construct applications that run on a variety of platforms, including iOS, Android, and the web. The static type system in Dart can aid in error detection at the compilation stage, resulting in fewer defects and a more dependable app. Generally speaking, Dart is a solid option for app development due to its combination of performance, productivity, flexibility, and powerful typing.

For Authentication and Database, Firebase has been used. Google created the Backend-as-a-Service (BaaS) platform called Firebase. It offers a variety of tools and services to developers so they can quickly and easily create and run mobile and web applications without having to deal with complicated infrastructure. Realtime databases, cloud Firestore, authentication, cloud storage, cloud functions, hosting, and other functionalities are all included in Firebase. Developers frequently utilize it to create scalable and dependable mobile and online applications.

## 4 AMITY - HYBRID APPLICATION

Amity is a Hybrid Application that works on all platforms such as Android, iOS, and the Web. It aims to provide assistance in all aspects of wellness that is Mentally, Socially, and Physically. It majorly focuses on Mental health wellness and brings forth an NLP-based chatbot for the end users who require support for their limited time anxiety, depression, and other sort of Mental issues. In need of immediate support, the end user is connected to medical assistance. The NLP Chatbot "DAZAI" converses with the end user as a companion and understands the user's problems and gives them therapeutic responses. The AMITY CONSTELLATION is the chat group where the end users can create, search, and join various groups of their interests and share their thoughts, ideas, and knowledge to help out the other peers. Common diet and

Exercise plans for various illnesses concerning mental health have been curated and posted in the Recommendation service. Qualified Doctors who are a part of the Amity Community are listed with their details in a separate section along with the Profile Information of the individuals.

## 4.1 OVERALL ARCHITECTURE

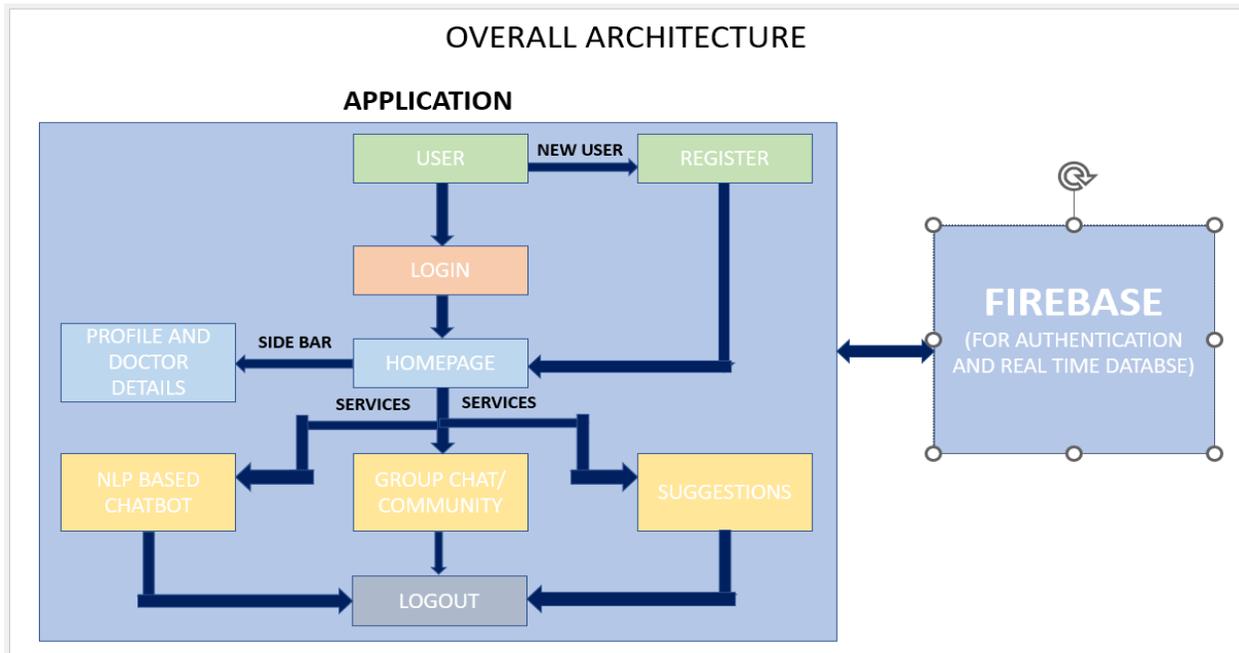

**Fig. 1.** Overall Architecture Diagram of the Amity Application.

Figure 1 shows the overall architecture of the AMITY Application. The user operates the application via their respective smart device. A new user is directed to the registration page, where the new user fills up the registration form. After successful registration, the user is directed to the Home Page. If the user is already registered he is directed to the login page, similarly after successful login, the user is directed to the home page. The Authentication and data are saved in the Firebase database which is connected to the Application via the backend. On the Home page, the sidebar provides user profile details and doctor details who are associated with AMITY. The main page consists of three services or functionalities. The first one is NLP Chatbot DAZAI for immediate assistance. The second one is the Online Group chat community AMITY CONSTELLATION which is a chat group for peer support. The last one is the suggestions page for diet and exercise curated by the Qualified dietitians associated with AMITY. There is also a Logout option in the sidebar which is used to come out of the application once the user is finished using the services.

## 4.2 FUNCTIONALITIES

There are two main features of the application we put forth.Firstly,an NLP based Chatbot DAZAI is designed to provide a companion-like feel to the user and provides therapy.The second vital Functionality is the community Chat group named Amity Constellation developed using Flutter Framework, Dart Programming Language, and Firebase for backend.. Constellation aims to connect users with similar problems so that they can help each other and feel better. It provides a medium for many who are users recovering from some kind of trauma to gain confidence and explore newer things that are suggested by users who might have faced a similar issue.Other Minor Functionalities involve Doctor Details, a Profile Page, and recommendations for Diet and Exercise to the users.

**4.3.1 NLP CHATBOT - DAZAI**

**A) PREPROCESSING DATA**

The JSON file containing 72 samples and 3 attributes is retrieved using the Python library pandas and is stored in a data frame. The dimension of the data frame becomes 246x3. The head of the dataset is displayed in Table 1. Using the tokenizer object, words in the sentence are split and converted to lowercase. Next, the text patterns are converted to sequences of integer values where each integer corresponds to a specific word in the vocabulary built by the tokenizer. Padding is done to sequences to make all the sequence lengths the same. The example of the dataset is shown in Table 1.

| tag | patterns | responses |
| --- | --- | --- |
| greeting | [Hi, Hey, Is anyone there?, Hi there, Hello, Hola…. | [Hello there. Tell me how are you feeling today.. |
| morning | [Good morning | [Good morning. I hope you had a good night's sleep. |
| afternoon | [Good afternoon | [Good afternoon. How is your day going?] |
| fact-26 | [What do I do if I'm worried about my mental health | [The most important thing is to talk to someone you trust. This might be a friend, colleague, family member, or GP. In addition to talking to someone, it may be useful to find out more information about what you are experiencing. |

|  |  | These things may help to get some perspective on what you are experiencing, and be the start of getting help] |
|---|---|---|
| Anger Management | [I easily recognize this but have no control over it and need suggestions for managing anger,....] | [I suggest that you work on your emotional awareness…..] |

Table. 1. A part of the mental therapy conversation dataset.

**B) MODEL ARCHITECTURE**

We use Long short-term memory Networks (LSTM) since they can effectively handle sequential data such as text.LSTMs remember long-term dependencies by using memory cells and can handle sequences of variable length.

The following layers are involved in the model:

i)Input layer- This layer defines the input shape of the model which is the maximum sequence length.

ii)Embedding(input_dim=vocab_size+1,output_dim=100,mask_zero=True) - This layer performs the embedding of words and converts the integer encoded-word sequences into vectors.

iii)LSTM(32, return_sequences=True) - This layer is a long short-term memory layer with 32 units- It passes the sequence as output to the next layer.

iv)LayerNormalisation()-Normalisation of outputs from the previous layer is performed.

v)Dense(128, activation="relu") - A dense fully connected layer is
deployed followed by a dropout layer. A softmax function is used at the last layer.

The model trains for 25 epochs with 98.78 accuracies. The testing accuracy is calculated by dividing the number of correct responses by the total number of responses which is 66%(20/30).

**C) INTEGRATION**

We use the Flask web application framework for creating applications in Python.The Chat application DAZAI gets input from the user and sends it to the machine learning model which is hosted on the local server. The model processes the input and responds with an appropriate sentence which is sent back to the chat interface.

### 4.3.2 GROUP CHAT

### A) ARCHITECTURE

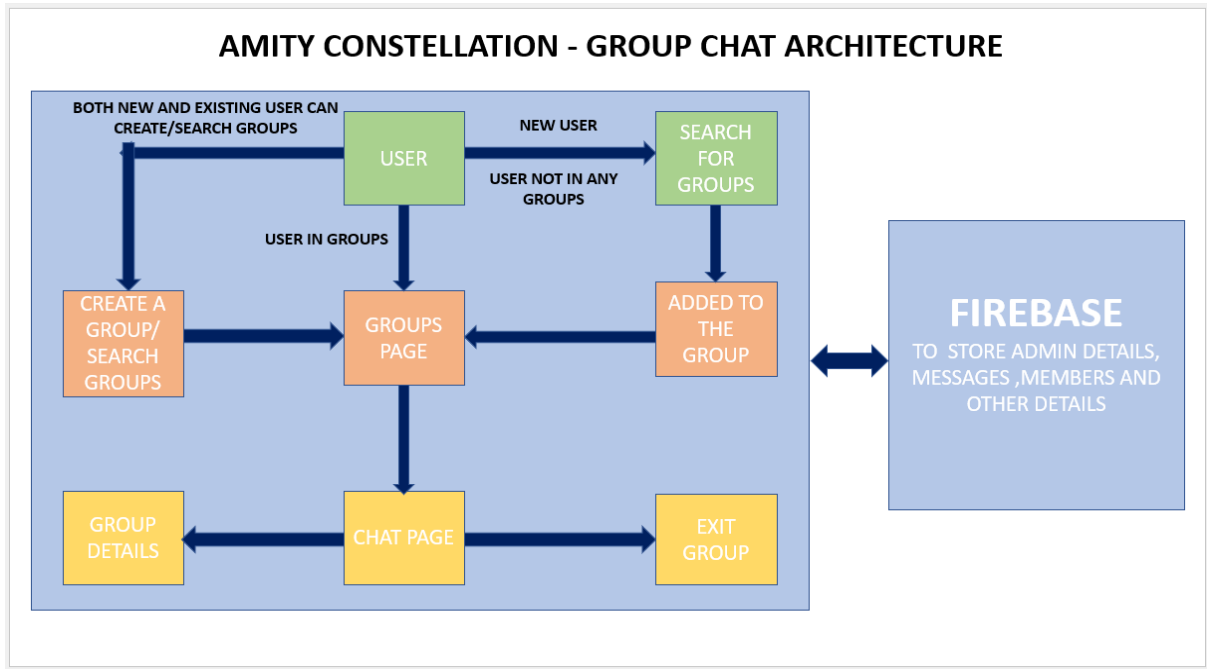

**Fig. 2.** Architecture Diagram of the AMITY CONSTELLATION - The Community Group.

Figure 2 shows the architecture of the Community Chat Group. Users can search for groups and join them. There is a Create group option for all users. All users can create and join any group. Immediately they are directed to the Groups page which shows the groups that the user is part of. On tapping the Group Icon, it is directed to the Chat page where the users can see the chat messages sent by other peers existing in the group and also they can post up their messages via this Page. On the Chat Page, there are two other functionalities which are, group details and Exit Group. The Group Details Page shows the member details along with admin for reference. The exit option allows the user to leave the group if the user is not interested.

### B) DEVELOPMENT

The Group chat is designed using Dart and is connected to Firebase. Firebase saves the group names, admin names, members, and the messages sent and received for every user. For Searching a Group a Query runs on the Firebase and then the desired group is shown according to the user's query. Then on pressing the join query, the user is added to that group. The maximum number of members that could be in a group is 256. For each group, a Firebase database is created with a unique group id and every time a message is sent a query runs and updates the database. The create, update, search, and delete queries are executed. The

Member details function retrieves the data from the Firebase with details of the members and the Admin.

### 4.3.3 OTHER FEATURES

The primary key used is the email id for every user created. The authentication process is done by Firebase with a mail id and password. The main page has a Profile page with basic information about the user. There is a page dedicated to showcasing details of doctors as well with information such as name, description, timings, and address of their clinic and contact number. The suggestions page has diet and exercise plans for concerns such as anxiety, depression, and hypertension.The Logout Function once pressed makes way for the user to log out of the application.

## 5 OUTCOME

In the phase after the development, integration, and execution we did a few testing of the execution of our Application. In Figure 3 below the NLP Chatbot - "DAZAI" is in full action. The user is in the process of talking with Dazai about their anxiety. Dazai replies with a therapist-based answer of practicing mindfulness-based activities like breathing techniques to the user.

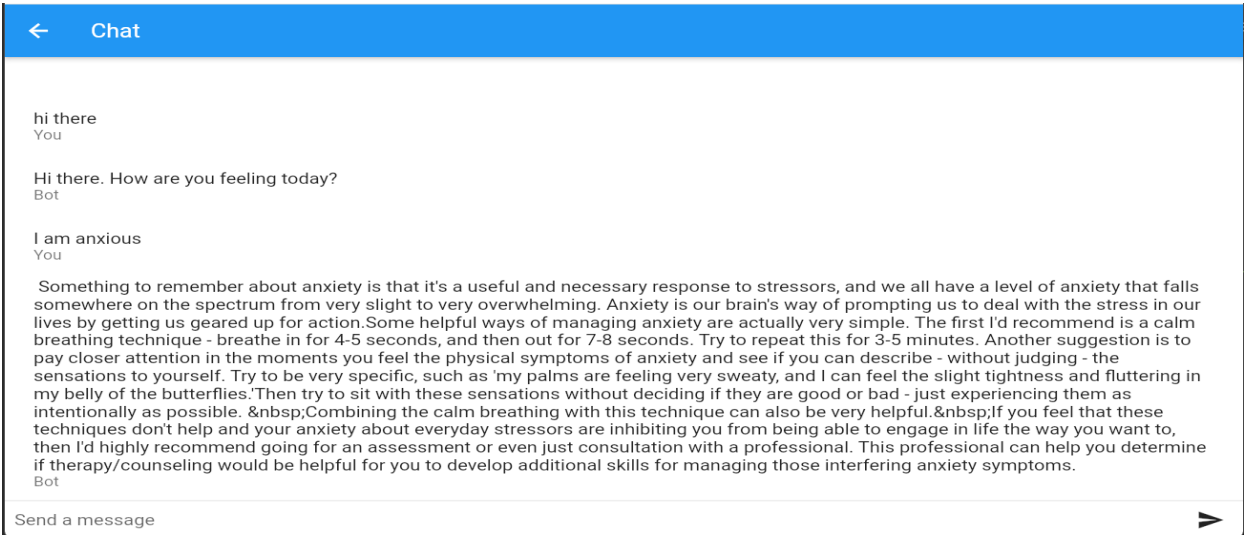

**Fig. 3.** The NLP Chatbot - DAZAI.

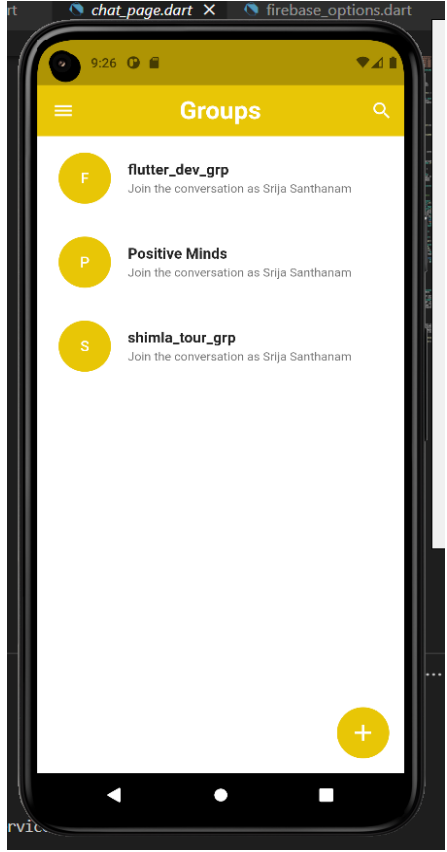

**Fig. 4.** AMITY CONSTELLATION - The Community Group - Group Page.

The Figure 4 is the Groups page of the application where a few groups of which a user is a part of is shown in the page.

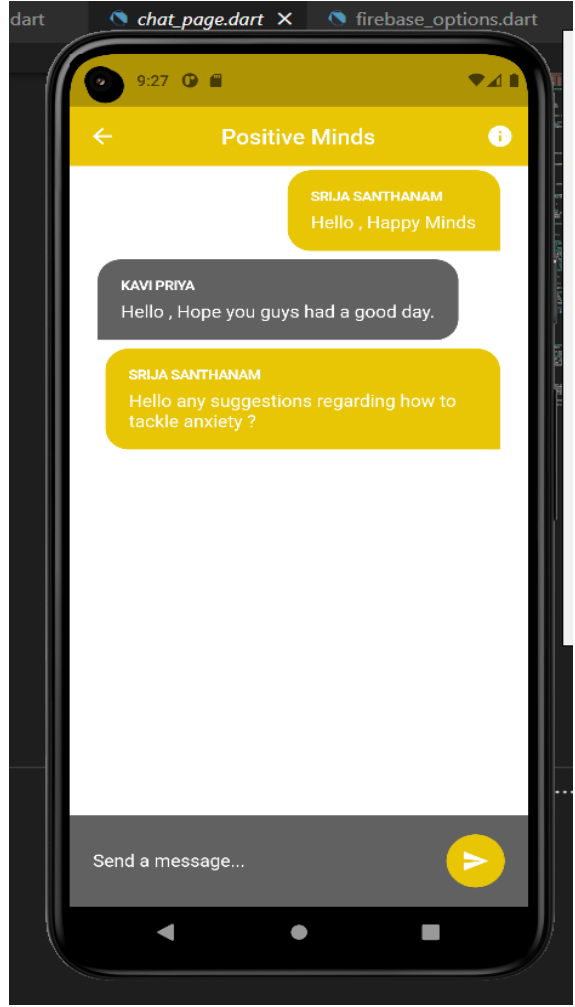

**Fig. 5.** AMITY CONSTELLATION - The Community Group - Chat Page.

The Figure 5 is the Chat page of the application where the users can chat with peers.

**6 DISCUSSION**

Usability testing is a sort of user testing that focuses on assessing a product's usability and user-friendliness, such as a website, mobile app, or software program. It entails watching and analyzing how actual users interact with the product to find any usability difficulties, challenges, or development opportunities. It includes collecting input from users as they execute activities or situations in a controlled setting and then using that feedback to make educated design decisions and improve the overall user experience. This is an important aspect of the user-centered design process because it ensures that a product is simple to use, efficient, and fits the demands of its intended consumers.

For our Application Amity, Around 30 individuals from various backgrounds had the experience of using our application. We divided our parameters into four major parts which are attractiveness, efficiency, dependability, and novelty. We asked the users to give a point on a

10-point Likert scale for each of the parameters individually and we took the average of the data and posted it as a graph in Figure 6.

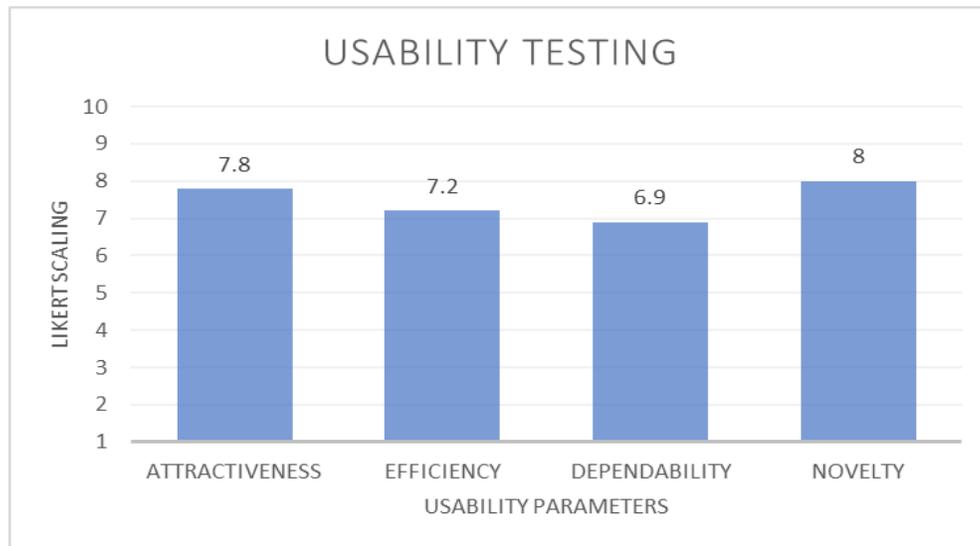

**Fig. 6.** Usability Testing outcome graph.

We conducted the Usability Tested and plotted the graph where on average the application has 78% of good attractiveness in terms of UI, users gave an 72% heads up for efficiency, 69% for dependability which is most likely to use the application again, and around 80 % for the novelty and uniqueness for the Amity Application.

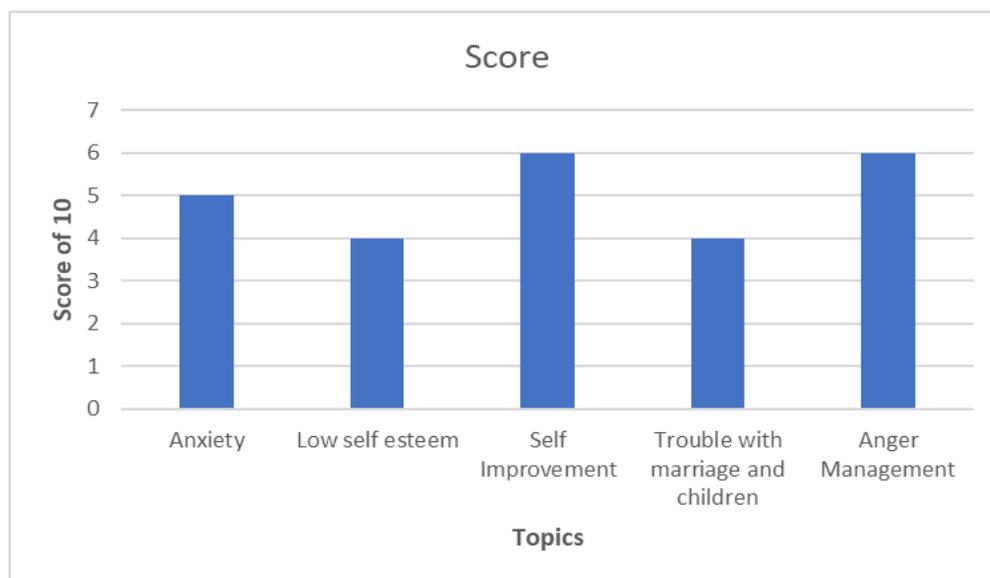

**Fig. 7.** Graph of the Score of the Tags Used

The tags in the dataset contain the above-mentioned major topics along with a few others. Each topic is allotted with 10 questions and is tested with the model.The results are plotted in Fig.7..The model is not equally trained concerning all the situations.

## 7 CONCLUSION AND FUTURE WORK

In this paper, we created a Hybrid Application - Amity which is used for mental help, where we developed an NLP Chatbot named Dazai. Dazai acts as a companion and provides therapist-like answers to the user's problems. The community chat group created as Amity Constellation is for every individual to interact with their peers and discuss mental health issues. The suggestions Page curated by the Medical team brings the user a sense of assistance. While the application overall looks appealing, the chatbot is quite inaccurate and must be trained on a lot of scenarios. The dataset we used is small and needs to be expanded. The model can be improved to a better accuracy. The app in general has got good understanding and acceptance from the users who were a part of the usability testing. In the future, the Application should be developed as a full-fledged all-service application with the inclusion of Appointment booking, Personalized Suggestions, Doctor's Feed, and more.

## 8 ACKNOWLEDGEMENTS

We extend our Thanks to the Mental Health therapists, counselors, and Doctors who helped us to extend our analysis and basic understanding of the topics concerning psychology. We would also like to thank the users and the members who were part of our usability testing survey.